\documentclass[twoside,twocolumn,9pt]{article}
\usepackage{extsizes}
\usepackage[super,sort&compress,comma]{natbib} 
\usepackage[version=3]{mhchem}
\usepackage[left=1.5cm, right=1.5cm, top=1.785cm, bottom=2.0cm]{geometry}
\usepackage{balance}
\usepackage{mathptmx}
\usepackage{sectsty}
\usepackage{graphicx} 
\usepackage{lastpage}
\usepackage{booktabs}
\usepackage[format=plain,justification=justified,singlelinecheck=false,font={stretch=1.125,small,sf},labelfont=bf,labelsep=space]{caption}
\usepackage{float}
\usepackage{fancyhdr}
\usepackage{fnpos}
\usepackage{comment} 
\usepackage[english]{babel}
\addto{\captionsenglish}{%
  
}
\usepackage{array}
\usepackage{droidsans}
\usepackage{charter}
\usepackage[T1]{fontenc}
\usepackage[usenames,dvipsnames]{xcolor}
\usepackage{setspace}
\usepackage[compact]{titlesec}
\usepackage{hyperref}

\usepackage{epstopdf}

\definecolor{cream}{RGB}{222,217,201}

\begin{document}


\makeFNbottom
\makeatletter
\renewcommand\LARGE{\@setfontsize\LARGE{15pt}{17}}
\renewcommand\Large{\@setfontsize\Large{12pt}{14}}
\renewcommand\large{\@setfontsize\large{10pt}{12}}
\renewcommand\footnotesize{\@setfontsize\footnotesize{7pt}{10}}
\makeatother

\renewcommand{\thefootnote}{\fnsymbol{footnote}}
\renewcommand\footnoterule{\vspace*{1pt}%
\color{cream}\hrule width 3.5in height 0.4pt \color{black}\vspace*{5pt}} 
\setcounter{secnumdepth}{5}

\makeatletter 
\renewcommand\@biblabel[1]{#1}            
\renewcommand\@makefntext[1]%
{\noindent\makebox[0pt][r]{\@thefnmark\,}#1}
\makeatother 
\renewcommand{\figurename}{\small{Fig.}~}
\sectionfont{\sffamily\Large}
\subsectionfont{\normalsize}
\subsubsectionfont{\bf}
\setstretch{1.125} 
\setlength{\skip\footins}{0.8cm}
\setlength{\footnotesep}{0.25cm}
\setlength{\jot}{10pt}
\titlespacing*{\section}{0pt}{4pt}{4pt}
\titlespacing*{\subsection}{0pt}{15pt}{1pt}

\renewcommand{\headrulewidth}{0pt} 
\renewcommand{\footrulewidth}{0pt}
\setlength{\arrayrulewidth}{1pt}
\setlength{\columnsep}{6.5mm}
\setlength\bibsep{1pt}

\makeatletter 
\newlength{\figrulesep} 
\setlength{\figrulesep}{0.5\textfloatsep} 

\newcommand{\topfigrule}{\vspace*{-1pt}%
\noindent{\color{cream}\rule[-\figrulesep]{\columnwidth}{1.5pt}} }

\newcommand{\botfigrule}{\vspace*{-2pt}%
\noindent{\color{cream}\rule[\figrulesep]{\columnwidth}{1.5pt}} }

\newcommand{\dblfigrule}{\vspace*{-1pt}%
\noindent{\color{cream}\rule[-\figrulesep]{\textwidth}{1.5pt}} }

\makeatother

\author{Varghese Babu,\textit{$^{a}{^\ddag}$}
Dilip Gersappe\textit{$^{a}$}}
\twocolumn[
\begin{@twocolumnfalse}

\title{Numerical model for ice formation in soil samples} 
\maketitle

{
	\centering
}
\abstract{Ice formation, or in general liquid crystallization, in porous media is important in various engineering contexts ranging 
        from frost heave in soils to structural damage due to salt crystallization. Developing mesoscopic models for this problem is challenging due to the multitude of phenomena involved - fluid flow, phase transitions and complex porous geometries. In this work, we present an efficient numerical implementation addressing this class of problems. We use a phase field model with a non-conserved order parameter for liquid crystallization and solve it in a Lattice-Boltzmann framework. Our modifications to the phase field model enable the direct capture of important thermodynamic phenomena such as the Gibbs-Thomson effect and Soil Freezing Characteristic Curve (SFCC). We demonstrate our model's capabilities in sphere packings as well as in realistic soil geometries. Our model is able to resolve the pore-scale freezing and capture the macroscopic freezing behavior such as the hysteresis observed during freezing and thawing.} \\
\end{@twocolumnfalse}
]

\renewcommand*\rmdefault{bch}\normalfont\upshape
\rmfamily
\section*{}


\footnotetext{\textit{$^{a}$~State University of New York, 100 Nicolls Rd, Stony Brook, 11794, NY, USA 
$^{\ddag}$~E-mail: varghese.babu@stonybrook.edu}}




\section{Introduction}

\label{intro}

Understanding pore-scale ice formation in soil is of great 
importance as it governs the microstructure, hydraulic 
permeability and susceptibility of soils to frost heave.
Frost heave, where soil 
deforms upward during freezing, causes substantial economic and structural damage and therefore requires a fundamental understanding \cite{dashPhysicsPremeltedIce2006} of the underlying mechanisms of ice formation.
Contrary to intuition, the expansion of water as it 
freezes
is not the sole reason for frost heave 
\cite{taber1930mechanics}. When ice forms in 
frost susceptible soil, it forms as large 
crystals in a phenomenon known as ice segregation \cite{schollickSegregatedIceGrowth2016,xiaFrostHeaveStudies2006,torranceXrayComputedTomography2008}. 
Ice segregation and the resulting frost heave are generally understood to occur in fine, silty soil, although exceptions exist \cite{tengFrostHeaveCoarsegrained2023}.
Finer soils possess small pores that result in a freezing 
point depression described by Gibbs-Thomson relation \cite{hagen@calpolyGibbsThomsonEquation2016,scalfiGibbsThomsonEquation2021b}. Therefore, when cooling the 
system below the liquid's bulk freezing temperature, only a portion 
of the pores freeze as dictated by local pore 
geometry \cite{ginotSoluteStronglyImpacts2020}. This process is coupled with another phenomenon called cryosuction which involves migration of liquid water 
within the system to the ice crystal front. 
\cite{guidaMicrostructuralInsightEffect2026}. This 
suction is understood to be due to the pressure difference 
between water at local equilibrium near the ice-water interface and the pressure at the water reservoir 
\cite{styleGeneralizedClapeyronEquation2023,gerberStressAccumulationConfined2022}. 
Consequently, access 
to the water reservoir is necessary for frost heave. 

A pore-scale model for frost-heave therefore requires three ingredients 
i) ice-formation controlled by the pore geometry in accordance with Gibbs-Thomson 
condition; ii) transport of liquid water towards the growing freezing front; iii) 
the deformation of the soil skeleton due to the crystallization pressure. In this 
work we build a model to address the first condition, which forms the thermodynamic 
basis for the other two as the crystallization pressure is 
set by the interface curvature at the pore throat and drives cryosuction. 
It should be noted that the physics behind frost heave is general and applicable to any liquid freezing in porous media and as such 
represents a general and important class of 
problems. 
This arises in understanding the damage to infrastructure due to salt 
crystallization 
\cite{desarnaudPressureInducedSalt2016}, changes
the permeability of porous media 
\cite{godinhoDynamicNatureCrystal2016}, in phase change materials \cite{fengUnidirectionalFreezingPhase2015} and in Artificial Ground Freezing 
\cite{niggemannInfluenceFreezingDirections2024}.

Numerical modeling of 
crystallization in porous media is challenging as
it involves modeling of liquid transport through the 
porous media, phase transformation of 
the liquid and the interaction of the phases with 
the pore walls. In addition, the geometries that 
need to be modeled are usually irregular and in 
three dimensions, requiring complex meshing. See 
\cite{liStateoftheArtConstitutiveModelling2024,liStateArtCoupled2025} for a review of numerical approaches.
 
Recently, numerical models based 
on Lattice-Boltzmann Methods (LBM) have emerged, taking advantage of the inherent 
advantages of LBM in handling complex geometries \cite{songMacroscopicLatticeBoltzmann2017,xu2019model,wangThawdrivenPermeabilityEvolution2026,liWaterIcePhase2024}. For instance, Song et al \cite{songMacroscopicLatticeBoltzmann2017} uses a continuum homogenized representation of the soil matrix without explicitly resolving local pore geometry. On the other hand, Wang et al \cite{wangThawdrivenPermeabilityEvolution2026} and Li et al \cite{liWaterIcePhase2024} resolve the simulations at pore-scale, however their frameworks are in two dimensions and the Gibbs-Thomson relation is incorporated into the system by modifying the temperature field according to the pore-geometry to prevent freezing. A similar 
approach is taken by Daneshian et al \cite{daneshian2026pore} in 3D: each fluid node is assigned the radius of the largest inscribed sphere that contains it, which sets a local freezing temperature through the Gibbs-Thomson relation. A node freezes only when the local temperature is below this value and a neighboring node is already frozen.  
 
In addition to LBM, Phase Field Methods (PFM) have also been used in modeling ice growth \cite{tanModelingIceCrystal2022,demangePhaseFieldModel2017,galfrePhaseFieldMethod2023}. 
Phase field models for solidification have been coupled with fluid 
\cite{beckermannModelingMeltConvection1999,rojasPhasefieldlatticeBoltzmannMethod2015,ratkaiPhasefieldLatticeBoltzmann2019}.
However, studies which use phase field modeling to study crystallization in porous media 
are fewer \cite{strachotaThreedimensionalPhasefieldSimulations2025,sunPhaseFieldModeling2023,suhMultiphasefieldMicroporomechanicsModel2022}. Among these, Suh et al 
\cite{suhMultiphasefieldMicroporomechanicsModel2022} model the soil matrix as a continuum without 
resolving pore geometries. Sun et al \cite{sunPhaseFieldModeling2023} use a phase field model in 
two dimensions with the temperature field modified to capture the effect of Gibbs-Thomson relation. 
Strachota et al \cite{strachotaThreedimensionalPhasefieldSimulations2025} solve the phase field in 3D for a collection of mono-disperse sphere packing. 
These efforts are primarily concerned with quantitative prediction of the amount of 
unfrozen water in the system.

In this work we propose a new framework for studying solidification in 3D realistic porous media. We make use of recent developments in solving phase field model using a LBM framework \cite{cartaladeLatticeBoltzmannSimulations2016,fakhariPhasefieldModelingMethod2010,tan2024lattice}. We implement our phase field model in OpenLB \cite{olbPaper2021}, which  
enables us to leverage GPU acceleration and use a modification similar to volume-penalization-method to incorporate the effect of complex pore walls. This scheme 
overcomes a key challenge in LBM : while standard fluid flow LBM relies on bounce-back 
at curved boundaries, the phase field requires an imposition of Dirichlet boundary 
condition at the pore wall to properly minimize the free energy and recover curvature 
dependent Gibbs-Thomson relation. Our method successfully captures key 
thermodynamic phenomena associated with crystallization in porous media such as the Gibbs-Thomson 
relation, SFCC and the hysteresis observed during freeze-thaw cycle. We demonstrate that the model is capable of handling realistic soil geometries such as one obtained 
through X-ray computational tomography (CT) scans. 


\section{Methods}

In this section we describe the phase field model used and the Lattice-Boltzmann updates used to solve the differential equations. 

\subsection{Phase field equations for solidification}
\begin{figure}
	\includegraphics[width=1\linewidth]{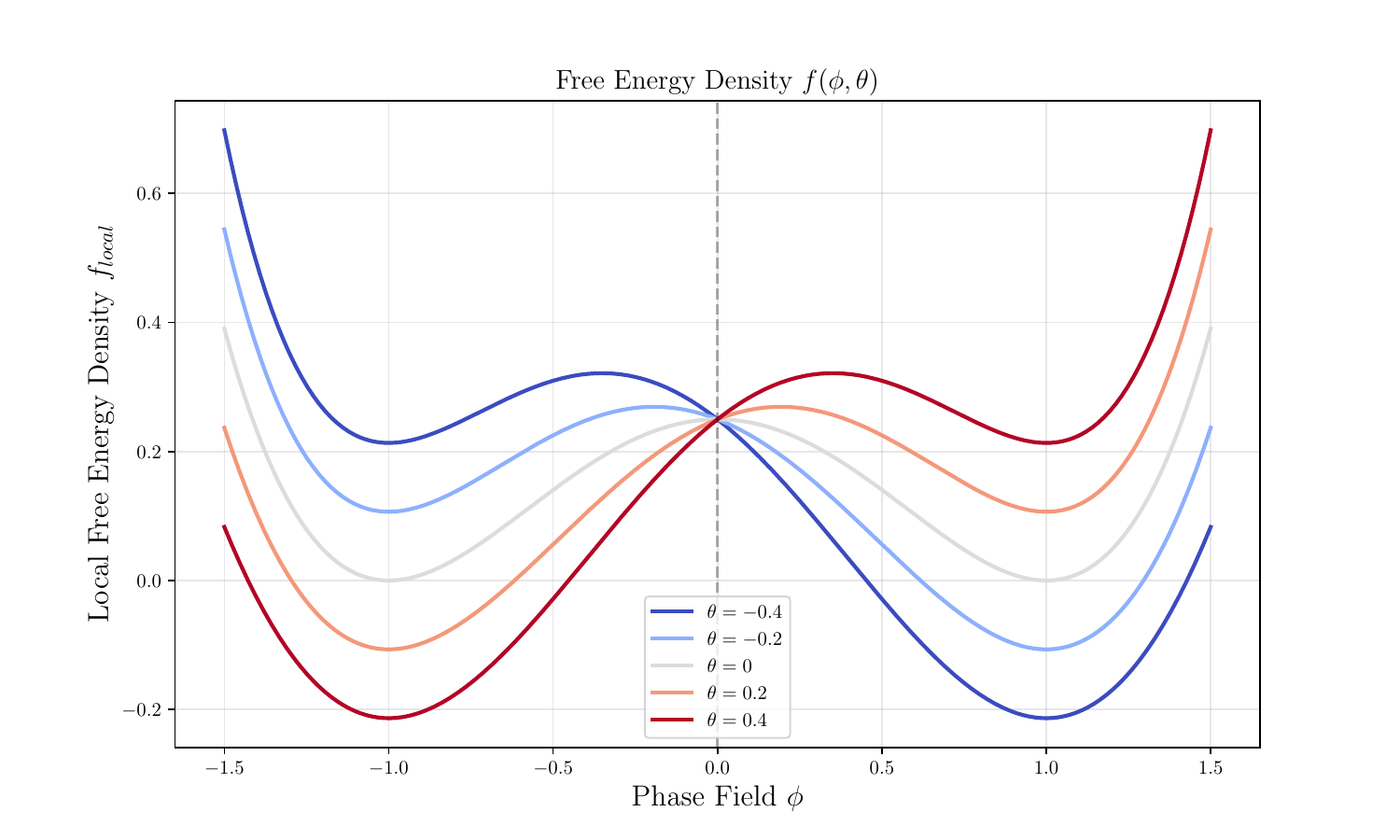}
	\caption{Local free energy density $f_{local}(\phi,\theta)=f_{dw}(\phi)+\lambda \theta G(\phi)$ for different values of $\theta$. Depending on the value of $\theta$, the system is biased to prefer one of the two minima.}
    \label{LocalFreeEnergy}

	\centering
\end{figure}
\label{methodPFM}
The phase field method is designed to overcome the difficulties in solving what are called the moving boundary problem 
or the Stefan problem. In problems such as solidification, the interface between the phases evolves over time. This poses difficulties
when solving the heat equation in both phases as the boundary conditions depend on the position of the interface, which requires solving the heat equation in turn. 
The phase field method overcomes the difficulty of tracking the interface between two phases by introducing a free energy functional with an order parameter $\mathcal{F} \left[\phi(\mathbf{x})\right]$ which takes different values in different phases. In the context of solidification, 
we conventionally use $\phi= \pm 1$ for the solid and liquid phases respectively. This 
order parameter can be thought of as a measure of the crystalline order or density 
difference. The evolution of the order parameter can then be deduced from a 
minimization of the free energy functional.  If the order parameter is not conserved, such as the one we are using here, then 
\begin{equation}
    \tau\frac{\partial\phi}{\partial t}=\frac{\delta\mathcal{F}}{\delta\phi}
\end{equation}
In this work, we start with the following free energy functional (divided by the free energy barrier density \cite{plappPhaseFieldModellingSolidification}),  
\begin{equation}\label{FreeEnergy}
\mathcal{F}(\phi,\theta)=\int\left(\frac{1}{2}W^2\left|\nabla\phi\right|^{2}+\overbrace{\frac{1}{4}(1-\phi^{2})^{2}}^{f_{dw}}+\lambda\theta G(\phi)\right)d\mathbf{x}
\end{equation}
Where $G(\phi)=\frac{1}{5}\phi^{5}-\frac{2}{3}\phi^{3}+\phi$ is an interpolation function whose
derivative is $G'(\phi)=(1-\phi^{2})^{2}$ and $f_{dw}$ is the double well potential. $\theta$ is the reduced dimensionless temperature, $\theta=\frac{C_{p}(T-T_{m})}{L}$, with $T_m$ the bulk freezing temperature, $C_p$ the specific heat and $L$ the latent heat of fusion, per mass. 
The term coupled to $\theta$ biases the local free energy density to 
either $+1$ or $-1$ depending on $\theta$, see Fig. \ref{LocalFreeEnergy}. 
The evolution equations for $\phi$ and $\theta$ are given by
\begin{align}
	\label{PFM}
	\tau\frac{\partial\phi}{\partial t} & =W^{2}\nabla^{2}\phi-\frac{\partial f_{dw}}{\partial\phi}-\lambda\theta G'(\phi)\\
    \label{HE}
	\frac{\partial\theta}{\partial t} & =\kappa\nabla^{2}\theta+\frac{1}{2}\frac{\partial\phi}{\partial t}
\end{align}.

We note that the equation for the evolution of temperature ($\theta$) cannot be 
derived from the above free energy functional, see \cite{karmaQuantitativePhasefieldModeling1998} for more details. 
There are 4 parameters that determine the behavior of the equation
- $(\tau,W^{2},\lambda,\kappa)$. $\tau$ has dimension of time, $W$
has dimensions of length, $\lambda$ is dimensionless and $\kappa=\frac{[L]^{2}}{[T]}$
is the heat diffusivity. The heat diffusivity $\kappa$ is related to thermal conductivity $k$ by $\kappa = \frac{k}{\rho C_p}$ where $\rho$ the density. 
An important feature we need to capture is the
Gibbs-Thomson condition  
\begin{equation}\label{GIBBS-THOMPSON}
	\theta = \frac{d_0}{R} - \beta V 
\end{equation}
where $d_0$ is the thermal capillary length and $\beta$ the kinetic coefficient, and $R$ is the curvature of the interface (in 2D). The thermal capillary length is given by 
\begin{equation}
\label{thermal_cap}
d_0 = \frac{k_g\gamma_{iw}}{\rho L} \frac{C_pT_m}{L} 
\end{equation}
where $\gamma_{iw}$ is the surface free energy of the ice-water interface and $k_g$ the Gibbs-Thomson coefficient which depends on the pore geometry.
We can relate $d_0$ and $\beta$ to model parameters \cite{karmaQuantitativePhasefieldModeling1998,plappPhaseFieldModellingSolidification} as 
\begin{equation}
    d_{0}=a_{1}\frac{W}{\lambda}
    \label{eq:d_0}
\end{equation}
and 
\begin{equation}
 \beta=a_{1}\left[\frac{\tau}{\lambda W}-a_{2}\frac{W}{\kappa}\right]   
 \label{eq:beta}
\end{equation}
$W$ is proportional to the interface thickness and dictates $d_0$ through Eq. \ref{eq:d_0}. Following similar studies in \cite{galfrePhaseFieldMethod2023,tan2024lattice,karmaQuantitativePhasefieldModeling1998}, we fix $\beta=0$ and therefore, 
\begin{equation}
\label{lambda}
    \lambda = \frac{\tau\kappa}{a_2W^2}
\end{equation}. 
$a_2$ and $a_1$ are constants that depend on the specific functions chosen for $f_{dw}$ and $G(\phi)$, and in our case $a_1=0.8838$ and $a_2=0.6267$.
Note that we can make heat diffusivity $\kappa$ to be a function of $\phi$ and $\mathbf{x}$ in Eq. \ref{HE}. 

\subsection{Using Lattice Boltzmann updates to solve Phase Field equations}
A recent advance in solving phase field equations is the use of Lattice-Boltzmann updates \cite{cartaladeLatticeBoltzmannSimulations2016,zhengLatticeBoltzmannModel2006,fakhariPhasefieldModelingMethod2010}. 
The standard LBM update equations are a discretization of the Boltzmann equation with the Bhatnagar–Gross–Krook (BGK) approximation. 
It can be shown using the Chapman-Enskog analysis that the macroscopic equations solved by the LBM equations are the 
Navier-Stokes equation and the continuity equation \cite{krugerLatticeBoltzmannMethod2017}.

Similarly, one can formulate LBM like discretization, a streaming step followed by a collision step, to solve a partial differential equation such as the phase field equation and heat equation \cite{cartaladeLatticeBoltzmannSimulations2016,younsiAnisotropyFunctionCrystal2016}. 
Chapman-Enskog analysis shows these equations solve the phase field and the heat equation \cite{cartaladeLatticeBoltzmannSimulations2016}. Similar to standard BGK-LBM, 
we introduce two lattices - one for phase field and one for heat equation. Each lattice has $n$ intermediate variables which depend on the exact lattice used. These variables, similar to the populations in standard LBM, evolve through a collision and streaming step. We use $g_i(\mathbf{x})$ for phase field variables and $f_i(\mathbf{x})$ for heat equation variables. 
The discretization equation for the $\phi$ field are, 
\[
	g_{i}(\mathbf{x}+\mathbf{e}_{i}d x,t+d t)=g_{i}(\mathbf{x},t)-\frac{1}{\eta_{\phi}}\left[g_{i}(\mathbf{x},t)-g_{i}^{eq}(\mathbf{x},t)\right]+\omega_{i}Q(\mathbf{x},t)\frac{d t}{\tau}
\]
where 

\[
	g_{i}^{eq}(\mathbf{x},t)=\omega_{i}\sum_{i=1}^{n}g_{i}(\mathbf{x},t)
\]
and
\[
	Q(\mathbf{x},t)=\left[\phi-\lambda\theta(1-\phi^{2})\right](1-\phi^{2})
\]
The value of $\phi$ at $(\mathbf{x},t)$ is given by $\phi(\mathbf{x},t)=\sum_i g_i(\mathbf{x},t)$.
Similarly for the heat equation $\theta$ we have,
\[
	f_{i}(\mathbf{x}+\mathbf{e}_{i}d x,t+d t)=f_{i}(\mathbf{x},t)-\frac{1}{\eta_{\theta}}\left[f_{i}(\mathbf{x},t)-f_{i}^{eq}(\mathbf{x},t)\right]+\omega_{i}Q_{\theta}(\mathbf{x},t)d t
\]
where 

\[
	f_{i}^{eq}(\mathbf{x},t)=\omega_{i}\sum_{i=1}^{n}f_{i}(\mathbf{x},t)
\]

\[
	Q_{\theta}=\frac{1}{2}\frac{\partial\phi}{\partial t}.
\]
and $\theta(\mathbf{x},t) = \sum_i f_i(\mathbf{x},t)$.
$\eta_\phi$ and $\eta_\theta$ are relaxation constants and are given by 
\begin{equation}
    \label{etaphi}
    \eta_\phi = \frac{1}{c_s^2} \frac{W^2}{\tau} \frac{dt}{dx^2}+\frac{1}{2}
\end{equation}
\begin{equation}
    \label{etatheta}
    \eta_\theta = \frac{1}{c_s^2} \kappa \frac{dt}{dx^2}+\frac{1}{2}
\end{equation}
where $c_s^2$ is the constant associated with the lattice. $dt$ and $dx$ are the time and space discretization, respectively. 
For the phase field equation we use a D3Q15 lattice and for the heat equation we use D3Q7 lattice. 

\subsection{Adding the effect of porosity to the Phase Field equations}

In designing the phase field model for porous media, the important feature we wish to capture is the Gibbs-Thomson effect (Eq. \ref{GIBBS-THOMPSON}). 
According to classical nucleation theory, the growth of a crystal in a 
super-cooled fluid is due to the competition between bulk free energy and 
surface free energy. For the crystal to grow, the decrease in free 
energy due to the bulk should compensate for the increase in free energy 
due to the surface that is created. Therefore, at low undercooling, 
small pores could exist in the system, inside which ice cannot form or 
enter, as the surface-to-volume ratio would be thermodynamically 
unfavorable for the crystal to grow. The Gibbs-Thomson relation is  
exploited to calculate the pore-size distribution in techniques like 
Cryoporometry and Thermoporometry \cite{hitchcockStudiesFreezingMelting2011,tolkkinenNMRCryoporometryPorous}.

The phase field model naturally captures the effect of interface energy due to the gradient term in Eq. \ref{PFM}, which is an energy penalty for creating a surface. 
The effect of the porous walls in this context will depend on the boundary conditions on $\phi$ on the porosity walls. In simple geometries (such as a flat surface) applying a Dirichlet 
boundary condition on the pore walls, with the value of $\phi$ set to $-1$ captures the Gibbs-Thomson effect quantitatively. However, it is challenging to apply a Dirichlet BC on the curved walls in Lattice-Boltzmann method. On the other hand, in the case of FEM, 
meshing complex 3D geometries to apply Dirichlet BC on curved walls is also numerically challenging. To overcome these difficulties, we modify the phase field 
free energy to include the effect of porosity. 

We add a term to Eq. \ref{FreeEnergy} that depends on the porosity of the sample, 
\begin{equation}\label{ModifiedFreeEnergy}
	\mathcal{F}(\phi,\theta)=\int\left(\frac{1}{2}W^{2}\left|\nabla\phi\right|^{2}+f_{dw}+{P}\lambda\theta G(\phi)+{(1-P)}\, \zeta\,(1+\phi)^2\right)d\mathbf{x}
\end{equation}
with $\zeta$ as a constant.
The above modification ensures that in the void space ($P=1$), this equation reduces to the phase field free energy Eq. \ref{PFM}. 
Inside the porous media, $P=0$ and the second extra term ensures that $\phi$ variable doesn't change from $\phi=-1$. 
The contribution to the free energy from porous media, with $P$ = 0 and $\phi$ enforced to $-1$, is a constant and doesn't change with time. 
Therefore the system will try to minimize the interface area of the system,  enabling us to capture the Gibbs-Thomson equation in the presence of porous media. We note that this is similar to volume-of-solid \cite{tsetoglouVolumeofsolidImplicitVolume2025} or volume penalty method \cite{mauryNumericalAnalysisFinite2009}.

\subsection{System Sizes}
\label{systemsizes}
In the previous section we related thermal capillary length to the model parameters 
such that a physical system such as water can be simulated using the phase field 
model. In this section we highlight the challenges in performing a quantitative 
simulation. For various analysis in this work, we use the sphere packing provided 
by Suh et al \cite{suhPoreMorphologyBasedEstimationFreezing2024}. This is a poly-disperse packing whose dimensions are $150 \mu \text{m}\times 150 \mu \text{m}\times 150 \mu \text{m}$.

First we compute the value of $d_0$, the thermal capillary length.
When we put the following values, which correspond to water, in the expression for $d_0$ in Eq. \ref{thermal_cap},
\begin{align*}
	\gamma_{iw} & =3\times10^{-2}\,\text{J}\text{m}^{-2}\\
    \rho &= 910 \, \text{kg}/\text{m}^3\\
	C_{p} & = 4184 \,\text{J}\,\text{kg}^{-1}\text{K}^{-1}\\
	T_{m} & =273.15\,\text{K}\\
	L & =3.34\times10^{5}\,\text{J}\text{kg}^{-1} \\ 
    k_g & = 4  
\end{align*}
we get $d_0 = 1.35 \times 10^{-9} \text{m}$. Therefore, the physical capillary 
length $d_0$ is on the order of nanometers. For numerical stability, the interface width has be resolved in the LBM simulation by at least 2-3 lattice units. If the physical interface width is enforced, the required grid spacing would be $dx \sim \mathcal{O}({10^{-9}})\text{m}$. Modern computational capabilities of 
LBM are capable of handling $\sim\mathcal{O}(10^9)$ or $(1000^3)$ cells.  
Therefore, enforcing a strict physical interface width will allow us to study a system size of upto $1\mu \text{m}; (10^{-6}\text{m})$, which is not sufficient for studying bulk soil conditions. Therefore, to study domain of linear dimension $150\mu \text{m}$, our numerical $d_{numerical}$ has to be  at-least 100 times larger than $d_0$. 
Since, our objective is to study the equilibrium shape of the solid phase in the 
presence of pore walls, we can rescale the dimensionless temperature such that we 
get the desired curvature. Looking at Eq. \ref{GIBBS-THOMPSON}, 
\begin{equation}
    \label{rescaling_temperature}
  \frac{\theta_{numerical}}{d_{numerical}}=\frac{\theta}{d_0}  
\end{equation}
for the same 
curvature. Therefore, if we chose 
an undercooling $\Delta T=0.01K$, then we simulate the system at 
$\frac{d_{numerical}}{d_0} \Delta T$. As we show below, this is closely related to 
the critical nucleus size in the system. Consequently, this affects the dynamics 
of the simulation unless $\kappa$ is divided by $s=\frac{d_{numerical}}{d_0}$. Subsequently, $\tau$ also has to be modified (Eq. \ref{lambda})) and effectively results in $dt \xrightarrow{}s\times dt$ from Eq. \ref{etaphi} and Eq. \ref{etatheta}.

\subsubsection{Critical Nucleation Size}

From Eq. \ref{FreeEnergy}, we can calculate the critical nucleation size at a given 
undercooling $\theta$. The difference between the local free energy between the two 
phases is given by $$\Delta f = \lambda \theta \left[ G(1)-G(-1)\right]=\frac{16}{15}\lambda \theta. $$ The shape of the interface for a 1D planar solution is given (for $\theta=0$)
by $$\phi_0(x) = \tanh \left[\frac{x}{\sqrt{2}W} \right].$$ The interface energy 
can be calculated by integrating across the interface. Therefore, 
\begin{equation}
    \sigma = \int _{-\infty}^{\infty }\left[ \frac{W^2}{2}|\nabla \phi_0|^2+f_{dw}(\phi_0) \right] dx  = \frac{2\sqrt{2}}{3}W 
\end{equation} is the interface free energy. 
So if we consider a solid phase as a sphere of radius $R$ in liquid phase, then the net free energy change is given by 
\begin{equation}
    \Delta F = \Delta f \times \frac{4}{3}\pi R^3 -\sigma \times 4\pi R^2.
\end{equation}
Therefore, for the free energy to decrease, 
\begin{equation}
\label{criticalRadius}
R>R_c=\frac{2\sigma}{\Delta f}=\frac{5\sqrt{2}}{4}\frac{W}{\lambda \theta}.     
\end{equation}
Writing this as Gibbs-Thomson relation 
$R_c = \frac{2d_{numerical}}{\theta}$ gives $$d_{numerical}=\frac{5\sqrt2}{8}\frac{W}{\lambda} = a_1\frac{W}{\lambda}$$ which is same as Eq. \ref{eq:d_0}. 
Therefore, if we have to resolve the critical nucleus by a few ($\sim 6$) lattice 
points and the interface width is $W=3dx$, then from Eq. \ref{criticalRadius}, $\theta = \frac{a_1}{\lambda}$ is the lowest dimensionless undercooling possible. Therefore, the 
lowest undercooling we can study is constrained by  $d_{numerical}/d_0$ ratio and 
the fact that we need to resolve the critical nuclei by a few lattice nodes. 
For example, with $\lambda=6$, then $\theta \approx 0.16$. Note that this is non-
dimensional ``numerical'' temperature, which will have the curvature that corresponds to undercooling of $d_{0}/d_{numerical}\times \theta$. For example, if $dx=150\mu \text{m}/1024$ then, $T_{min}=272.9K$ is the lowest temperature we can simulate which can be expected to respect the curvature allowed by Gibbs-Thomson effect for parameters that correspond to water. If $dx=150\mu \text{m}/256$ then, $T_{min}\approx 273.1K$. 

Therefore, the lowest temperature we can simulate is dictated by the 
resolution. 

\subsubsection{Timestep}
However, increasing the resolution (i.e $dx$ is reduced) brings up another 
challenge. 
As the resolution $dx$ and timestep $dt$ are related to each other by Eq. \ref{etaphi} and Eq. \ref{etatheta}, when resolution is increased $dt$ has to be reduced quadratically. For example, let us fix $\eta_{\theta}=1.0$, which is a reasonable stable value for LBM, then 
\begin{equation}
    dt =  \frac{1}{2} dx^2 \frac{c_s^2}{\kappa}
\end{equation}

For ice, the thermal diffusivity is $\kappa = \frac{\alpha_{ICE}}{\rho C_p}$ with 
thermal conductivity $\alpha_{ICE}\approx 2.2 \,\frac{\text{W}}{\text{K}.\text{m}}$. If we simulate a cube of 150$\mu m$ in length, where we 
resolve each side with $1024$ lattices, then $dx=1.5\times 10^{-7}\,\text{m}$ and 
$dt=5\times 10^{-9}\, \text{s}$ from Eq. \ref{etatheta}. For ice to propagate from 
the top to bottom of a 150$\mu m$, it takes $\sim 2\, \text{s}$ (see Section \ref{unifreezesection}). Therefore, it takes a  large number of timesteps for the system to freeze over, if we resolve 
the system using a reasonable resolution. 

This  highlights the trade-offs in using this model for mesoscopic studies. We need to increase the 
resolution to study pore-scale ice formation, which the LBM based solver on GPUs can handle, 
but we pay the price by reducing the timestep which leads to longer 
simulations. 

{{We want to also clarify that despite the above constraints, the 
timescales and length scales relevant to the problem are beyond the 
capabilities of most other methods, even coarse grained Molecular Dynamics \cite{molinero2009water}. For example, given that a water 
molecule is about an few angstroms in size, therefore to simulate a micrometer (size 
of a silt particle) the number of particle we need would be a few hundred 
billion. In addition, the largest time scale that we can run MD simulations is often at the nanosecond scale, which is orders of magnitude smaller than our case.  We note that the constraints associated numerical modeling of this problem has to do with the fact that the relevant length-scale in the problem is the size of the smallest pore, which is inherent to the problem. We could study lower temperatures if we focus on ice formation on individual pores or a selected pore network where our resolution would be smaller.}}




\section{Results}

In this section we discuss the results from our implementation of the phase field model in porous media.

\subsection{Validation of Lattice Boltzmann as a technique to solve the phase field equations}
\label{FEMvalidationSection}
\begin{figure}
	\centering
	\includegraphics[width=\linewidth]{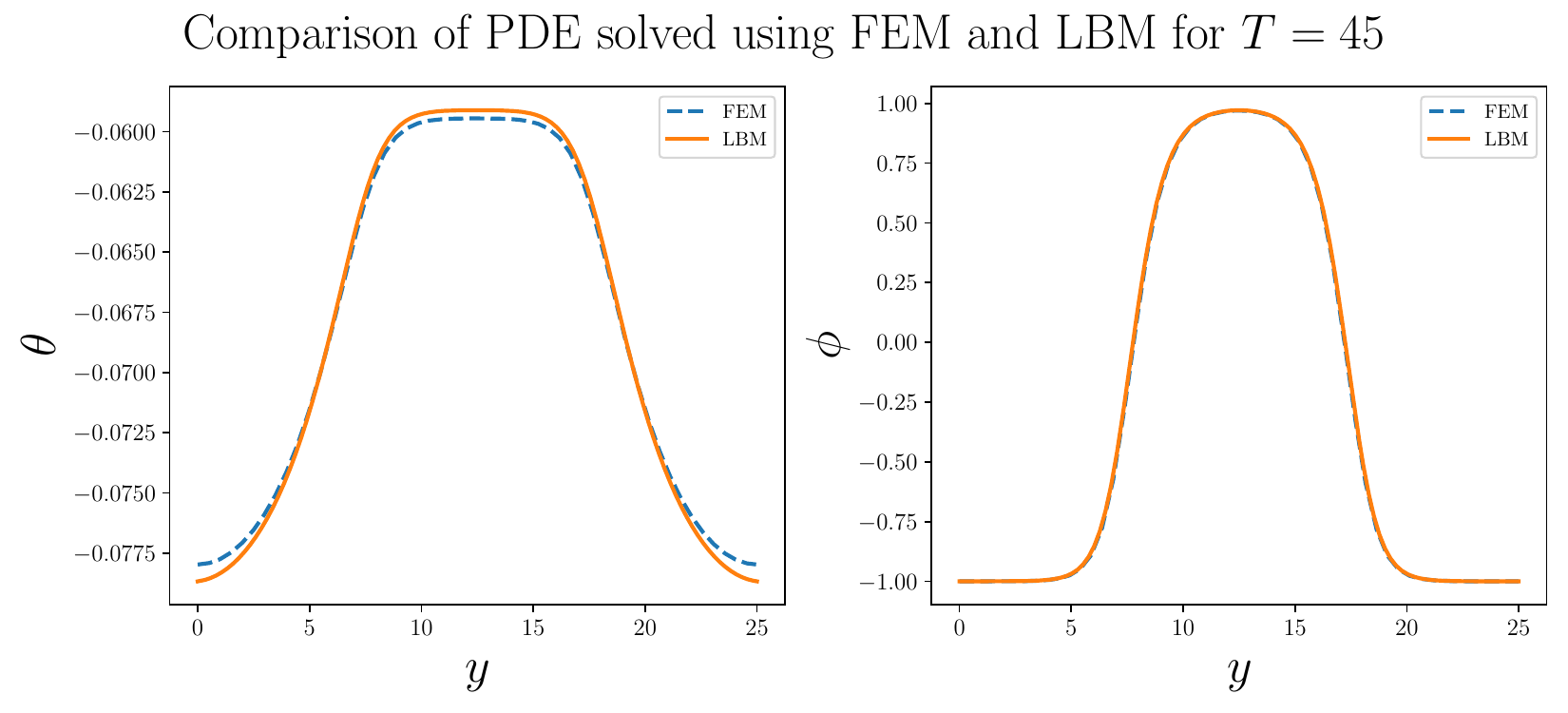}
	\caption{Comparison of FEM and LBM for solving the phase field equation of a solid nucleus growing in an undercooling.}
    \label{FEMValidation}
\end{figure}
In this section, we validate our Lattice Boltzmann method to solve the phase field model against standard 
techniques like FEM implemented in MOOSE. 
Similar to the benchmarks in \cite{wuPhaseFieldBenchmark2021b}, we initialize a solid phase at the center of a 3D domain. The solid phase is initialized as 
\begin{equation}
    \label{tanh}
\phi(x) = \tanh\left(\frac{R_s-|\mathbf{x}-\mathbf{x_0}|}{\sqrt{2W^2}}\right)
\end{equation}
We use $W=1; \tau =1; \kappa=4; \lambda =6.382$, $dx=0.25$, $dt=10^{-3}$ and an undercooling $\theta = 0.1$. 
The coupled equations with these parameters is solved by the LBM implementation in OpenLB and a FEM implementation in MOOSE and the results are shown in Fig. \ref{FEMValidation}. Our results show agreement between both computational tools.

\subsection{Unidirectional Freezing}
\label{unifreezesection}
\begin{figure}
	\centering
	\includegraphics[width=\linewidth]{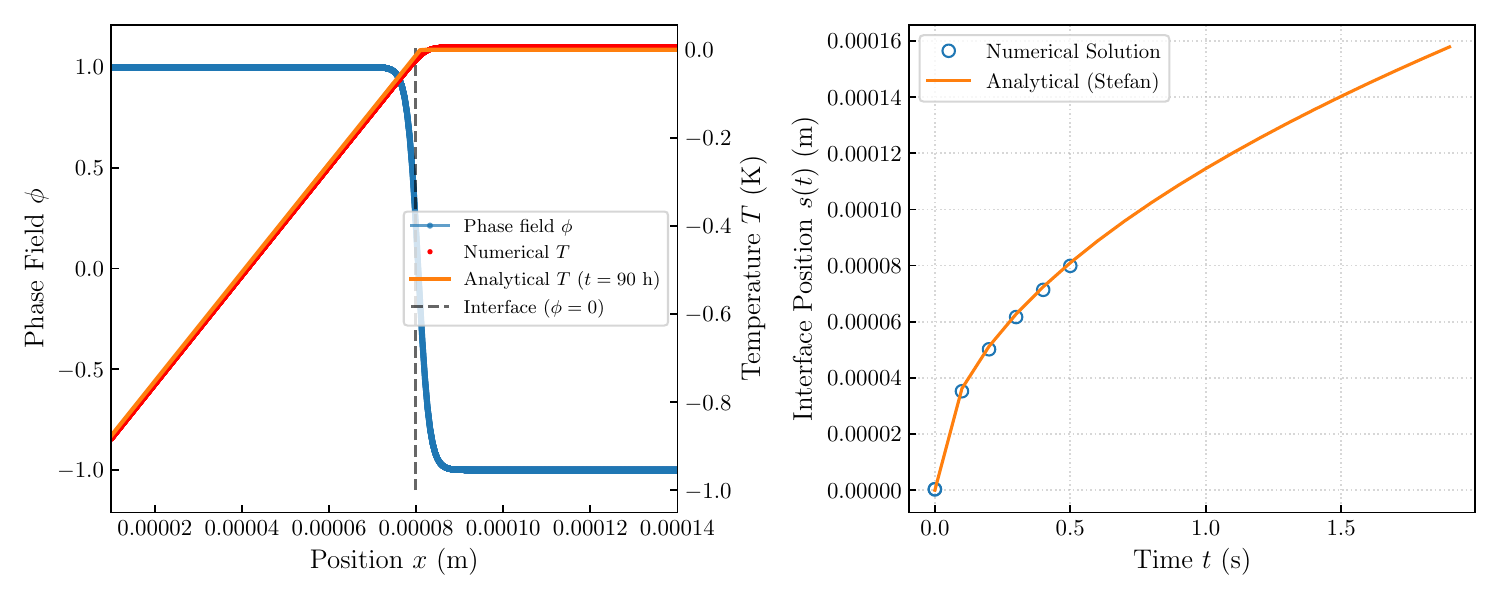}
	\caption{\label{UnifreezeFig} Validation of phase field method with the analytical solution of the 1D Stefan Problem.}
\end{figure}
\begin{table}[h]
	\centering
	\begin{tabular}{|cc|}
		\hline
		\textbf{Quantity} & \textbf{Values}\tabularnewline
		\hline
        \hline
		$L_{z}$ & $0.15\times 10^{-3} \, \text{m} = 150\mu \text{m}$\tabularnewline
		\hline
        $L_{x}=L_{y}$ & $0.025L_{z}$\tabularnewline
		\hline
        $dx$ & $L_{z}/256$\tabularnewline
        \hline
		$W$ & $3\times dx$\tabularnewline
		\hline
        $k_l$ & $0.58 \,\text{Kg.m.s}^{-3}\text{K}^{-1}$\tabularnewline
        \hline
        $k_s$ & $2.22 \,\text{Kg.m.s}^{-3}\text{K}^{-1}$\tabularnewline
        \hline
        $\rho$ & 1000 $\text{Kg.m}^{-3}$\tabularnewline
		\hline
        $C_p$ & 4184 $\text{m}^{2}.\text{K}^{-1}.\text{s}^{-2}$ \tabularnewline
		\hline
		$\lambda$ & $10$\tabularnewline
		\hline
		$\tau$ & $\frac{\lambda a_{2}W^{2}}{(\kappa_l+\kappa_s)/2}$\tabularnewline
		\hline
        $\eta_{\theta}$ & 1. \tabularnewline
		\hline
		$dt$ & $c_s^2\left(\eta_{\theta}-\frac{1}{2}\right)\frac{dx^2}{\kappa}$\tabularnewline
		\hline
		$a_{1}$ & $0.8838$\tabularnewline
		\hline
		$a_{2}$ & $0.6267$\tabularnewline
        \hline 
        $T_s$ & -1$^o$C \tabularnewline
        \hline 
        $L_{heat}$ & 334 kJ \tabularnewline
		\hline
	\end{tabular}
	\caption{\label{unifreezetable} \textbf{Uni-directional freezing} Summary of various values used to generate Fig. \ref{UnifreezeFig}.}
\end{table}
\begin{table}[h]
	\centering
	\begin{tabular}{|c|c|}
        \hline
		\textbf{Quantity} & \textbf{Values}\tabularnewline
        \hline
        \hline        
        $L_{x}=L_{y}=L_z$ & $0.15$mm\tabularnewline
        \hline
        $k_l$ & $0.58 \,\text{Kg.m.s}^{-3}\text{K}^{-1}$\tabularnewline
        \hline
        $k_s$ & $2.22 \,\text{Kg.m.s}^{-3}\text{K}^{-1}$\tabularnewline
        \hline
        $\rho$ & 1000 $\text{Kg.m}^{-3}$\tabularnewline
		\hline
        $C_p$ & 4184 $\text{m}^{2}.\text{K}^{-1}.\text{s}^{-2}$ \tabularnewline
		
        \hline
		$dx$ & $L_x/256$\tabularnewline
        \hline 
        $\eta_{\theta}$ & 1.0 \tabularnewline
        \hline
		$dt$ & $c_s^2\left(\eta_{\theta}-\frac{1}{2}\right)\frac{dx^2}{\kappa}$\tabularnewline
		\hline
		$W$ & 3$\times dx$ \tabularnewline
        \hline
		$\lambda$ & 6 \tabularnewline
		\hline
		$\tau$ & $\frac{\lambda a_{2}W^{2}}{\kappa_s}$\tabularnewline

		\hline
		$a_{1}$ & $0.8838$\tabularnewline
		\hline
		$a_{2}$ & $0.6267$\tabularnewline
        \hline 
        $d_{numerical}$ & 2.59$\times 10^{-7}$ m  \tabularnewline
		\hline
	\end{tabular}
	\caption{\label{gibbsthomsonparameters}\textbf{Gibbs-Thomson Effect} Summary of various values used to generate Fig. \ref{fig:gibbsthomson}.}
\end{table}

As discussed in section \ref{methodPFM}, phase field models are designed to solve the Stefan problem or the moving interface problem.
The Stefan problem of one dimensional freezing has an analytical solution and we can use that to validate our phase field method.
The solutions and their derivation can 
be found in \cite{kaaksEnergyconservativeDGFEMApproach2023,javierreComparisonNumericalModels2006}. The specific form of the analytical solution 
we use here is from \cite{kaaksEnergyconservativeDGFEMApproach2023}. 
The Stefan problem for 1D solidification is given by, 
\begin{align*}
\frac{\partial T}{\partial t} & =\kappa_{liquid}\frac{\partial^{2}T}{\partial x^{2}};0<x<X(t)\\
\frac{\partial T}{\partial t} & =\kappa_{solid}\frac{\partial^{2}T}{\partial x^{2}};X(t)<x<d\\
\rho_sL\frac{\partial X(t)}{\partial t} & =k_{solid}\frac{\partial T\left(X(t)\right)}{\partial x}-k_{liquid}\frac{\partial T\left(X(t)\right)}{\partial x}\\
T(0,t) & =T_{s};T_{s}<T_{m};\\
T(d,t) & =T_{m}
\end{align*}
where $X(t)$ is the interface position and $d$ is the length of the domain. One end at $x=d$ is kept at $T_m$ and at $x=0$ is kept at a $T_s<T_m$. $k$ is the thermal conductivity and $\kappa$, thermal diffusivity with $k=\rho C_p \kappa$.
The interface position is given with respect to time by 
$$X(t)=2\Lambda (\kappa_st)^{\frac{1}{2}}$$
where $\Lambda$ is the solution to the transcendental equation 
\begin{multline}
    \frac{\Lambda d \pi^{1/2}}{C_s[T_m - T(0,t)]} = 
    \frac{\exp\left(-\Lambda^2\right)}{\text{erf}(\Lambda)} \\
    - \frac{k_l}{k_s} 
    \frac{\kappa_s^{1/2}[T(x,0) - T_m]\exp\left(-\kappa_s \Lambda^2 / \kappa_l\right)}
    {\alpha_l^{1/2}[T_m - T(0,t)]\,\text{erfc}\!\left(\Lambda(k_s/k_l)^{1/2}\right)}.
\end{multline}
As shown in Fig. \ref{UnifreezeFig} we get very good agreement with 
the analytical result. In phase field models, usage of different 
thermal diffusivity of both phases has to be introduced carefully 
\cite{almgrenSecondOrderPhaseField1999,mcfaddenThinInterfaceAsymptoticsa}. This could be reason for the small difference in temperature profile between the numerical and analytical results, where the phase field model is smoothed out relative to the analytical solution at the interface. The values used to simulate unidirectional freezing is given in Table \ref{unifreezetable}. Note that this gives an idea about how long it takes for ice to propagate from one end to other for the system size we have chosen, and it is about 2s. 

\subsection{Effect of porosity}
\begin{figure}
	\includegraphics[width=\linewidth]{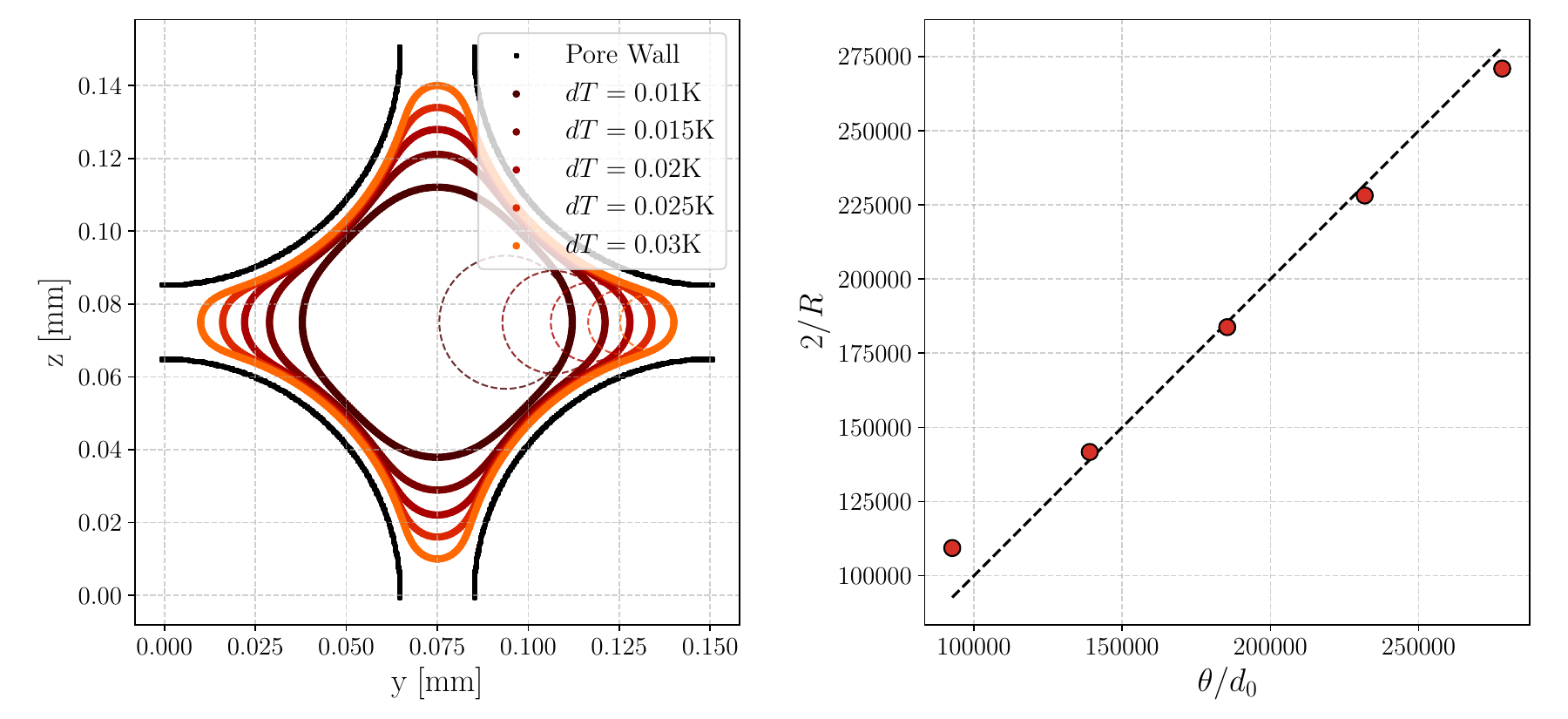}
	\caption{Validation of Gibbs-Thomson effect in the phase field model. \textbf{a)} Shows the contours of the phase field crystal ($\phi=0$) along with the pore wall
	for different temperatures. The curvature of the contour at the tips is measured by fitting a circle to the tip. The maximum curvature $\kappa=\frac{1}{R}$ of the solid phase is related to the 
	temperature by Eq. \ref{eq. gibbsthomson3D}. As shown in Fig \textbf{b}, the linear relationship holds for the temperatures studied. The dashed black line is Eq. \ref{eq. gibbsthomson3D}. This shows that the addition of porosity term to the free energy Eq. \ref{ModifiedFreeEnergy} 
	seems to capture the Gibbs-Thomson effect quantitatively. Note that we have used a $\theta_{numerical}$ which is the rescaled temperature as described in Section \ref{systemsizes}.}
    \label{fig:gibbsthomson}
	\centering
\end{figure}
As mentioned before, an important thermodynamic phenomenon that a model
for liquid freezing in porous media has to recreate is the Gibbs-Thomson relation. 
In $3D$, the Gibbs-Thomson effect can be written compactly as, 
\begin{equation}
	\label{eq. gibbsthomson3D}
	\theta = 2d_0\kappa
\end{equation}
where $\kappa$ is the mean curvature given by $\kappa = \frac{1}{2}\left(\frac{1}{R_1}+\frac{1}{R_2}\right)$ where $R_1$ and $R_2$ are the principal radii of curvature. 
For the symmetric case considered here we have $$\theta =\frac{2d_{0}}{R}.$$ 
To validate the modified free energy functional Eq. 
\ref{ModifiedFreeEnergy}, we test the phase field equation in an 
idealized pore geometry. We define a cubical domain with $8$ spheres 
placed at the corners. The spheres are represented in the domain by a 
porosity field $P$ which is $1$ in the void spaces and $0$ inside the 
spheres. A spherical nucleus of solid phase is initialized at the center of the 
domain using Eq. \ref{tanh}. The parameters are given in Table. 
\ref{gibbsthomsonparameters}. As discussed earlier, the interface width is chosen 
to be several orders of magnitude larger than the desired physical value. In this simulation, since we are only interested in the equilibrium shape of the phase field, we have switched-off the 
heat equation and the whole domain is held at the undercooling temperature. In addition, periodic boundary conditions are enforced in all directions. 

The solid phase is grown at different real undercoolings 
as shown in Fig. \ref{fig:gibbsthomson}. 
When this phase has reached equilibrium, we measure the curvature at the 
tip of the solid phase. Since the system is symmetric, we consider a  
plane parallel to a face of the cube and plot the contour of the 
solid phase defined by $\phi=0$ on the plane. To calculate the 
curvature, we find the coordinate of the tip and consider all data 
points within a search radius around that point. Then we fit those data points to a circle using least squares method. The radius calculated can then be used to calculate the curvature and verify 
Eq. \ref{eq. gibbsthomson3D}.

As seen in Fig. \ref{fig:gibbsthomson}, we get good agreement with 
expected behavior showing that the penalty term in Eq. \ref{ModifiedFreeEnergy} is able to enforce Gibbs-Thomson effect. 


\subsection{Soil Freezing Characteristic Curve (SFCC)}
\begin{figure}
    \centering
    \includegraphics[width=0.8\linewidth]{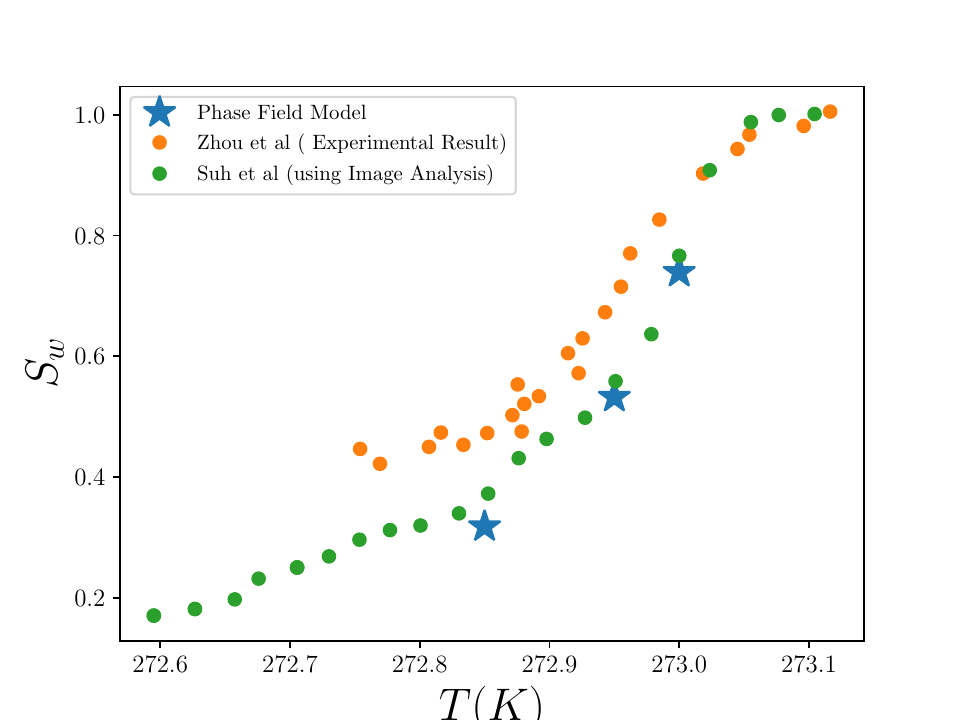}
    \caption{Amount of unfrozen water $S_w$ from numerical \cite{suhPoreMorphologyBasedEstimationFreezing2024} and experimental \cite{zhouSimultaneousMeasurementUnfrozen2014}, the values are in expected range, however further calibration is required to match the values.}
    \label{fig:unfrozenwaterfigure}
\end{figure}

\begin{figure}
    \centering
    \includegraphics[width=\linewidth]{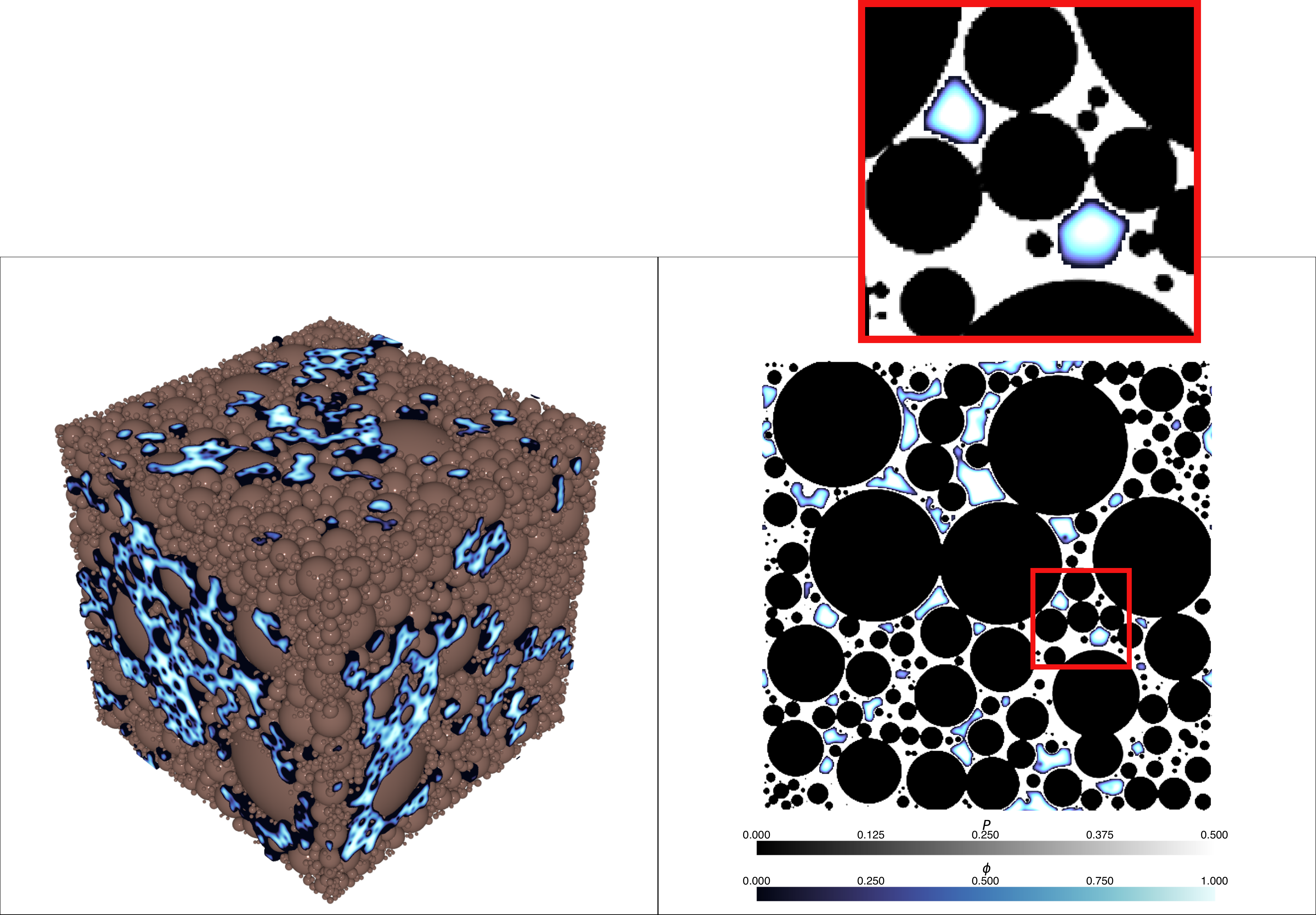}
    \includegraphics[width=\linewidth]{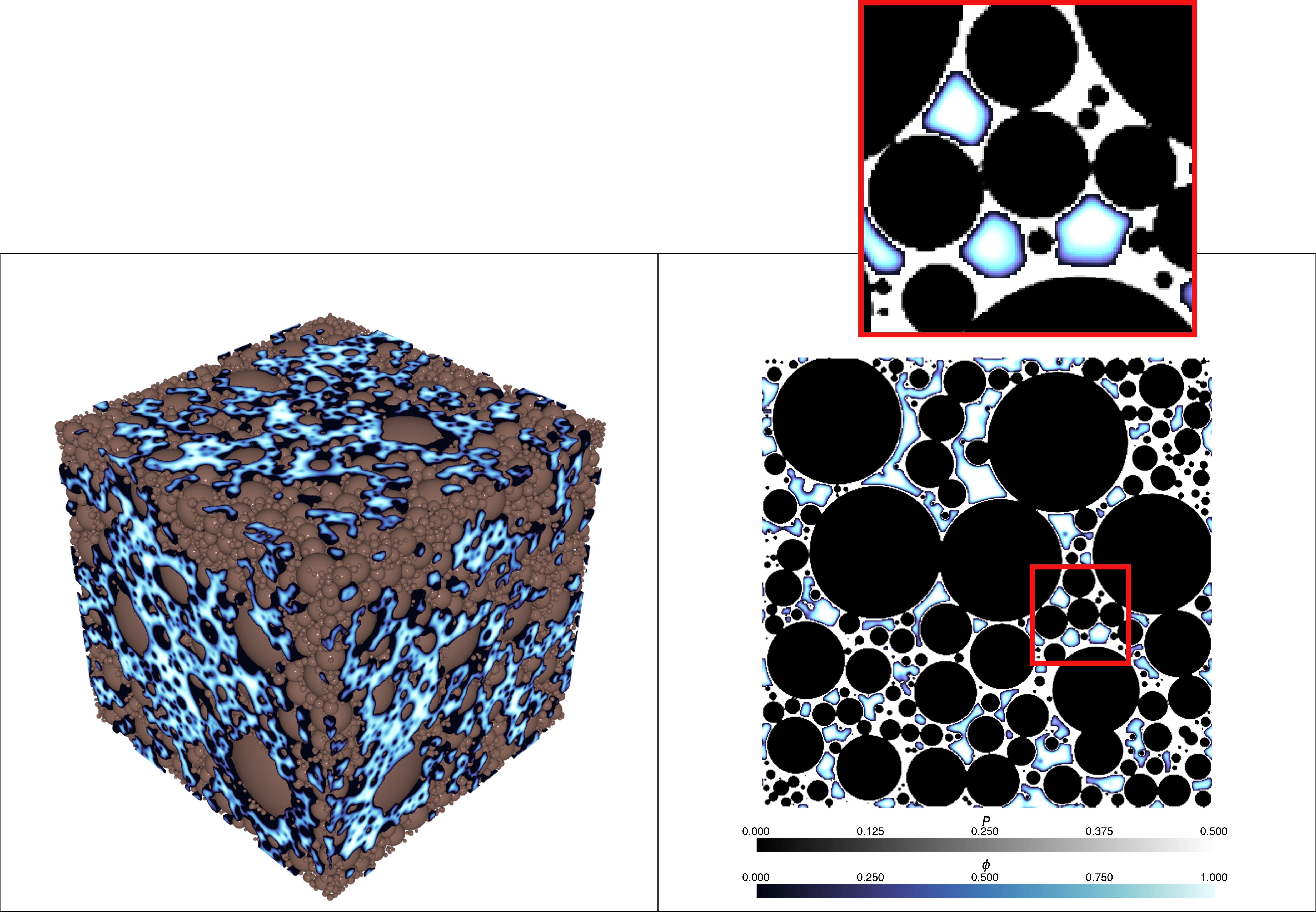}
    \includegraphics[width=\linewidth]{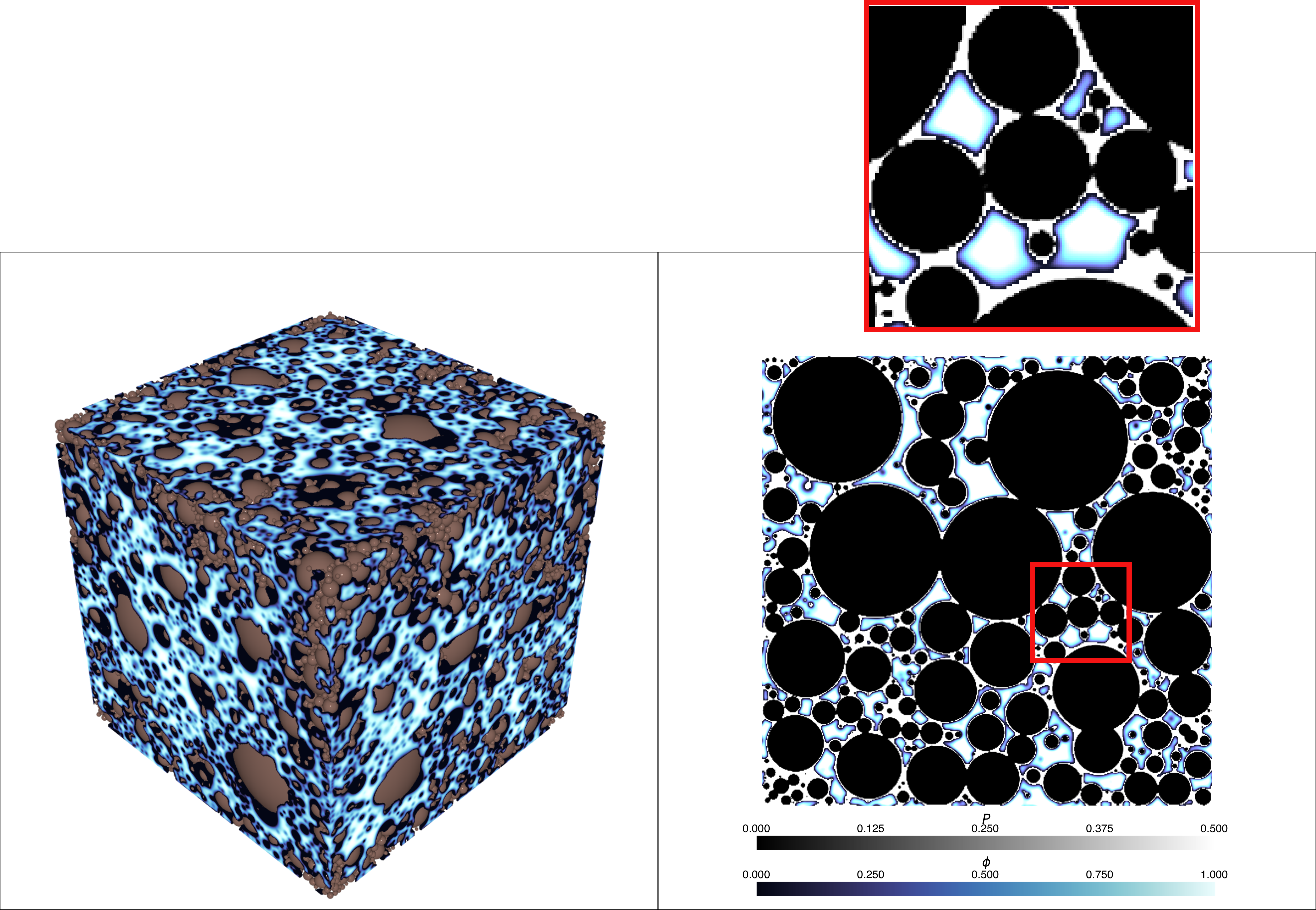}
    \caption{A snapshot of the simulation during freezing simulation. On the left is the 3D isosurface of the phase field $(\phi=0)$ showing the complex structure formed within the void spaces of the sphere packing. On the right is a 2D slice through origin parallel to a side of the cube. It shows how narrow channels formed by spheres in close contact stops the ice from propagating further. Figures from top to bottom corresponds to temperatures of $T=273K$, $T=272.95K$ and $T=272.85K$ with the insets at the top being a zoomed section highlighted using red square.}
    \label{fig:phaseFieldPolySpheres}
\end{figure}
Now that we have validated the phase field model for solidification, 
we can apply it to study liquid crystallization in realistic soil 
geometries. The most relevant feature of soil we are interested in is 
the particle size distribution. Particle sizes in soil are generally 
characterized by large size disparity, with a large number of fine particles 
along mixed a few much larger particles 
\cite{sunPhaseFieldModeling2023,suhPoreMorphologyBasedEstimationFreezing2024,zhouSimultaneousMeasurementUnfrozen2014}. For testing our system we use a sphere 
packing with a log-normal particle size distribution from Suh et al 
\cite{suhPoreMorphologyBasedEstimationFreezing2024}. The sample consists of a $N=15000$ spheres packed in a domain of 
$0.15\,\text{mm}\times 0.15\,\text{mm} \times 0.15\,\text{mm}$. 
We resolve this domain as $400\times 
400\times 400$ lattice units, therefore $dx=3.75\times 10^{-7}\text{m}$.
The rest of the parameters are the same as Table \ref{gibbsthomsonparameters}.
In this initial analysis, we are only interested in the amount of unfrozen water at 
equilibrium, therefore, we switch-off the heat equation for this case as well. 

We start with an initial condition with the whole domain as solid phase, $\phi=1$. As 
the simulation progresses, certain regions of the domain will be converted to the
liquid phase with $\phi=-1$ according to Gibbs-Thomson effect. The simulation 
results are visualized in Fig. \ref{fig:phaseFieldPolySpheres}. At low 
undercooling, many small pores in the system get converted to liquid. The 
configurations are at different temperatures and all from the same initial condition. 
Therefore, in the freeze-thaw cycle, this curve would trace the thaw values under
quasi-static conditions. At a given temperature, the solid phase forms in every pore 
which is larger than the critical size at that temperature. This may not match the 
experimental case where heterogeneous nucleation in pore walls is a stochastic 
process and therefore ice may not form in every pore possible. Thus, we can expect 
that the phase field methods as described above gives a lower bound for the amount 
of unfrozen water (maximum ice phase). 

The results are shown in Fig. \ref{fig:unfrozenwaterfigure}, where we have compared 
the values obtained from the phase field method with experimental values and the 
result of Suh et al \cite{suhPoreMorphologyBasedEstimationFreezing2024}. The amount of unfrozen water is given by 
\begin{equation}
    S_w = \frac{ \# \text{Nodes}(\{\phi<0)\})}{\# \text{Nodes} (\{ P > 0.5\})}
\end{equation} where $\# \text{Nodes}(\{\phi < 0\})$ is the number of nodes with $\phi < 0$ i.e water and is divided by number of nodes with $P>0.5$ i.e void space. We also calibrate the system by setting $\zeta=5.0$ in Eq. \ref{ModifiedFreeEnergy}. The values 
obtained from our method predicts a lower value of unfrozen water - perhaps due to 
the stochastic nature of heterogeneous nucleation in experimental systems. 

As mentioned, we use OpenLB to implement the phase field model in the Lattice-
Boltzmann framework. The simulations presented here are mostly run on NVIDIA RTX 
PRO 6000 Blackwell GPUs. Since we need to simulate two lattices, one for phase field and one for heat equation, our memory requirement is larger than that of standard LBM simulations. However, we observe very good performance. For results in 
this section, our lattice has $\sim 64$ million cells, and we achieve $\sim1730$ MLUPs ( Million Lattice Updates Per second).

\subsection{Hysteresis}


\begin{figure}
    \centering
    \includegraphics[width=\linewidth]{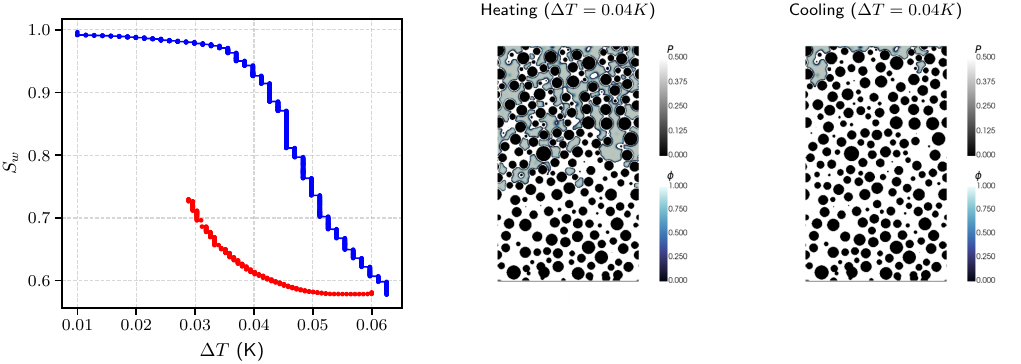}
    \caption{ Hysteresis in the amount of liquid phase during freezing and thawing. The staircase pattern occurs because of ice growth at given temperature. The two figures shows ice phase at the same temperature during the heating and the cooling phase.}
    \label{fig:Hysteresis}
\end{figure}

Similar to the previous section, the spheres are initialized as regions with porosity values equal to zero and the whole system is initialized with a temperature field of 
$T_{initial}=273.14 K$. Periodic boundary 
conditions are imposed in the $x$ and $y$ directions. The top and bottom 
boundaries are held at a constant temperature. 
We initialize a solid phase at the top of the 
domain to stimulate ice growth. We simulate ice growth till $\text{abs} \{ \frac{S_w^{i} - S_w^{i-1}}{S^{i}_w} \}<10^{-3}$, then we 
reduce the temperature by $dT=0.00142 \text{K}$ and 
repeat. As explained before, the constraints 
of the method limits us to very low 
undercoolings, therefore we only lower the 
temperature till where the critical 
nucleation radius given by Eq. \ref{criticalRadius}  
is greater than 2 lattice units. Here we don't switch off the latent heat as we did in the previous section, however we rescale the temperature to get the right curvature as per Eq. \ref{rescaling_temperature}. The sphere packing we use here is a cuboidal domain which is $0.3\,\text{mm}\times0.3\,\text{mm}\times0.5\,\text{mm}$. The system consists of 3000 spheres whose radii is a Gaussian distribution with mean radius of $1.12\times 10^{-2}\,\text{mm}$ and variance of $2.45\times 10^{-3}\,\text{mm}$. We use $dx=L_{z}/500$, $\eta_{\theta}=1$ and $dt$ given by Eq. \ref{etatheta}. The remaining values are same as the previous section. 

The result of freeze-thaw simulation is shown in Fig. \ref{fig:Hysteresis}.
An important feature of SFCC is the hysteresis observed during cooling 
and thawing. The observation of hysteresis can be understood in the following 
way. If ice nucleates in the smaller pore of a network where a small pore is 
connected by a narrow channel to a larger pore, then ice can only pass through 
the narrow pore if the system cooled to a lower temperature. Once the large 
pore is frozen, when the system is thawed, the ice around the narrow channel 
might melt, but the large pore will continue to have ice inside it giving rise to a 
substantially different profile 
during freezing and thawing. This show that Gibbs-Thomson effect along with 
the pore network of the sample is sufficient for observation of 
hysteresis.



%
\subsection{Application of the phase field to realistic soil geometries}
\label{STLfile}
\begin{figure}[t]
    \centering
    \includegraphics[width=0.49\linewidth]{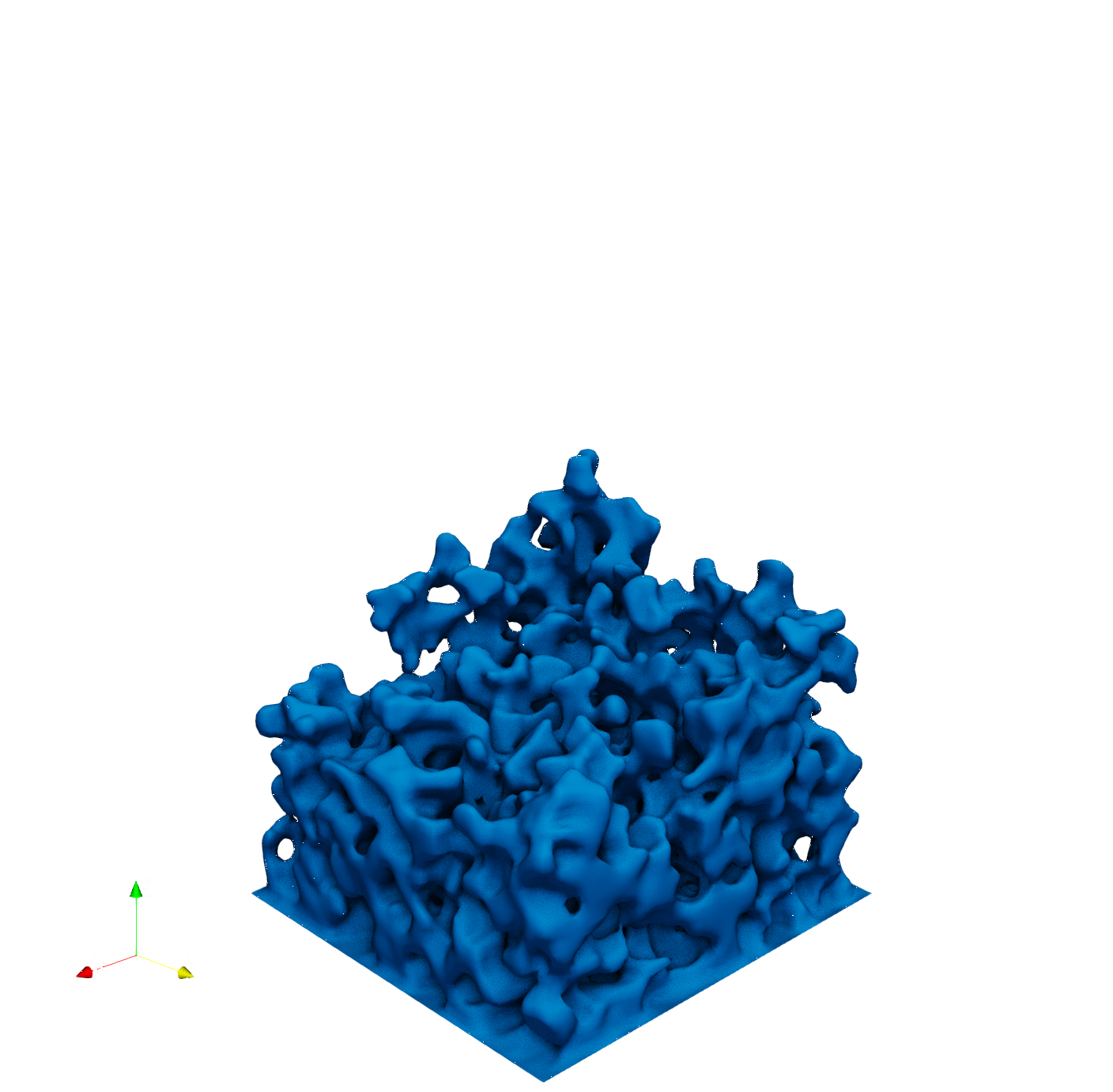}
    \includegraphics[width=0.49\linewidth]{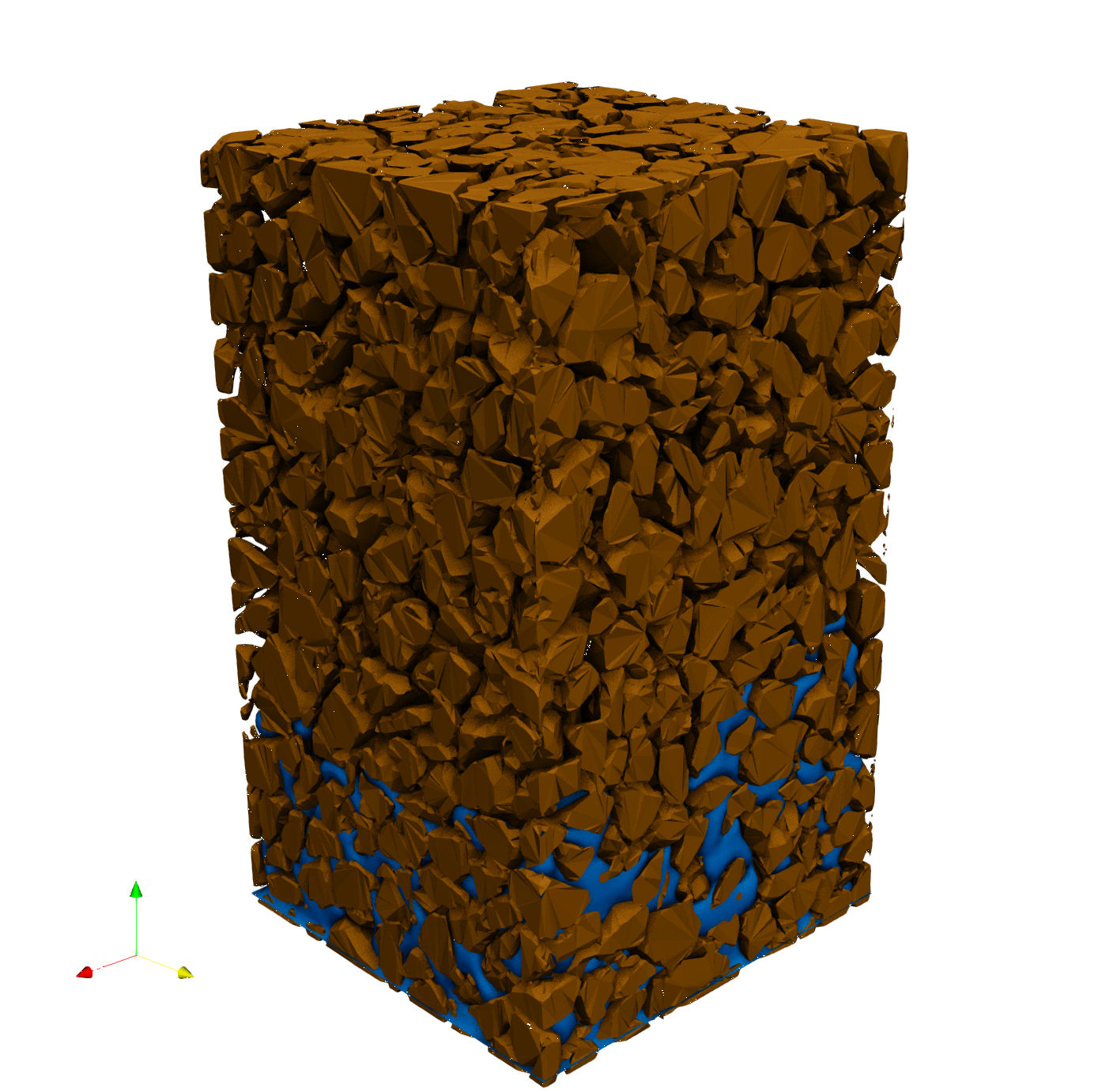}
    \caption{Left: Solid phase (isosurface $\phi=0$) growing in soil sample obtained through CT scan. Right: Shown along with the soil sample.}
    \label{fig:stlFile}
\end{figure}
\begin{table}[htpb]
    \centering
    \begin{tabular}{lccc}
        \toprule
        \textbf{Parameter} & \textbf{Symbol} & \textbf{Value} & \textbf{Unit} \\
        \midrule
        Heat diffusivity (water) & $\alpha_w$ & $1.0 \times 10^{-7}$ & m$^2$/s \\
        Heat diffusivity (ice) & $\alpha_i$ & $1.0 \times 10^{-6}$ & m$^2$/s \\
        Heat diffusivity (soil) & $\alpha_s$ & $1.5 \times 10^{-6}$ & m$^2$/s \\
        Volumetric heat capacity & $C$ & $4.23 \times 10^6$ & J/(m$^3 \cdot$ K) \\
        Volumetric latent heat & $L$ & $305.65 \times 10^6$ & J/m$^3$ \\
        Undercooling & $\Delta T$ & $9$ & K \\
        \midrule
        Grid spacing & $\Delta x$ & $0.0018$ & m \\
        Time step & $\Delta t$ & $200$ & s \\
        \bottomrule
    \end{tabular}
    \caption{Physical and numerical parameters used in the simulation shown in Fig. \ref{fig:stlFile}.    \label{tab:sim_parametersSTL}
}
\end{table}
We can use the phase field model developed above and apply it to soil geometry obtained through  a CT-scan. We use the publicly available dataset from \cite{nadimiMicroComputedTomography2024} 
for testing our method. We use the ``RoughSand Skeleton'' sample in the dataset for our calculation. The sample is rescaled to approximately $0.17\text{m}\times 0.17\text{m}\times 0.3\text{m}$ and other values are given in Table \ref{tab:sim_parametersSTL}. The visualizations from the simulation 
is shown in Fig. \ref{fig:stlFile}. Since, the phase field method can adapt to arbitrary porous media, we are not constrained to use spheres to represent the porous media. This simulation is only for demonstration and we do not present any quantitative analysis here. 


\section{Conclusions}

In this paper we have developed a phase field model solved 
through the Lattice Boltzmann frame work, to simulate liquid crystallization in porous 
media. We validated the model using analytical solution to Stefan problem and by comparing with finite element solution. A key advantage of the model is its 
ability to naturally recover the thermodynamically correct shape of the 
interface in constrained geometries as described by the Gibbs-Thomson 
relation. Our method does not require any geometric analysis on the porous structure to capture 
the Gibbs-Thomson effect as done in 
\cite{daneshian2026pore,sunPhaseFieldModeling2023}. 
We have shown that the 
model successfully captures macroscopic properties like hysteresis in soil 
freezing characteristic curve and the qualitative nature of SFCC. We have also highlighted the limitations of such a 
model, where resolving the nanopore-scale ice formation limits us to study of very low 
undercooling (or temperatures close to bulk freezing temperature) or systems of the 
order of $\mu \text{m}$.

Future work for the model has two primary 
directions. First one is incorporation and coupling the growth of solid phase 
with fluid flow building up on established phase field models 
\cite{beckermannModelingMeltConvection1999,tanModelingIceCrystal2022,ratkaiPhasefieldLatticeBoltzmann2019} to model cryosuction. 
The second direction is incorporating the crystallization pressure and 
the deformation of the porous medium.  
The OpenLB framework in which we have 
implemented this model can be coupled with DEM simulation frameworks like 
LAMMPS \cite{BABU2026615}. These multiphysics couplings will be pursued in future work. Overall, the model presented here provides the thermodynamic foundation and a  first step toward simulating frost heave in soils.

\section {Acknowledgments}
This project is supported by the United States Army Corps of Engineers, Engineer Research and Development Center’s Cold Regions Research and Engineering Laboratory (ERDC-CRREL, Contract No. W913E524C0009). Any opinions, findings, and conclusions or recommendations expressed in this material are those of the author(s) and do not necessarily reflect the views of the Broad Agency Announcement Program and ERDC-CRREL. The authors would like to thank Stony Brook Research Computing and Cyberinfrastructure and the Institute for Advanced Computational Science at Stony Brook University for access to the high-performance SeaWulf computing system, which was made possible by \$1.85M in grants from the National Science Foundation (awards 1531492 and 2215987) and matching funds from the Empire State Development’s Division of Science, Technology, and Innovation (NYSTAR) program (contract C210148).

\section*{Author contributions}
Varghese Babu: Methodology, Writing — original draft, review \& editing, investigation, visualization. Dilip Gersappe: Conceptualization, Supervision, Writing — review \& editing, Funding acquisition.

\section*{Conflicts of interest}
There are no conflicts to declare.

\section*{Data availability}
The code used for the simulations are available at 
\texttt{https://github.com/vavachan/PhaseFieldHeatEquation}.


\balance


\bibliography{crystallization.bib} 
\bibliographystyle{rsc} 

\end{document}